\documentclass[preprints,article,accept,moreauthors,pdftex]{mdpi} 

\firstpage{1} 
\pubvolume{xx}
\issuenum{1}
\articlenumber{5}
\pubyear{2020}
\copyrightyear{2020}
\history{Received: date; Accepted: date; Published: date}
\usepackage{amsfonts}
\usepackage{amsmath}
\usepackage{amssymb}

\pdfoutput=1

\newcommand{\lwk}{{{\rm low}\mbox{-}k}}
\newcommand{\vlwk}{$V_{{\rm low}\mbox{-}k}$}

\newcommand{\zbb}{$0\nu\beta\beta$}
\newcommand{\dbb}{$2\nu\beta\beta$}
\newcommand{\teff}{$\Theta_{\rm eff}$}

\newcommand{\heff}{$H_{\rm eff}$}

\newcommand{\nme}{$M^{0\nu}$}
\newcommand{\nmes}{$M^{0\nu}$s}
\newcommand{\nmed}{$M^{2\nu}$}

\newcommand{\nmeds}{$M^{2\nu}$s}

\Title{Present Status of Nuclear Shell-Model Calculations of
  $0\nu\beta\beta$ Decay Matrix Elements}

\Author{Luigi Coraggio $^{1,2}$*, Nunzio Itaco $^{1,2}$, Giovanni De
  Gregorio $^{1,2}$, Angela Gargano $^{2}$, Riccardo Mancino $^{1,2}$
  and Saori Pastore $^{3}$}

\AuthorNames{Luigi Coraggio, Giovanni De Gregorio, Angela
  Gargano, Nunzio Itaco, Riccardo Mancino and Saori Pastore}

\address{%
$^{1}$ \quad Dipartimento di Matematica e Fisica, Universit\`a degli
  Studi della Campania ``Luigi Vanvitelli'', viale Abramo Lincoln 5 -
  I-81100 Caserta, Italy\\
$^{2}$ \quad Istituto Nazionale di Fisica Nucleare, Complesso
Universitario di Monte  S. Angelo, Via Cintia - I-80126 Napoli,
Italy\\
$^{3}$ \quad Physics Department and McDonnell Center for the Space
Sciences at Washington University in St. Louis, St. Louis, Missouri
63130, USA}

\corres{Correspondence: luigi.coraggio@na.infn.it}

\abstract{
Neutrinoless double beta (\zbb) decay searches are currently among the
major foci of experimental physics.
The observation of such a decay will have important implications in
our understanding of the intrinsic nature of neutrinos and shed light
on the limitations of the Standard Model.
The rate of this process depends on both the unknown neutrino
effective mass and the nuclear matrix element (\nme) associated with
the given \zbb\,  transition.
The latter can only be provided by theoretical calculations, hence the
need of accurate theoretical predictions of \nme for the success of
the experimental programs. 
This need drives the theoretical nuclear physics community to 
provide the most reliable calculations of \nme.
Among the various computational models adopted to solve the many-body
nuclear problem, the shell model is widely considered as the basic
framework of the microscopic description of the nucleus.
Here, we review the most recent and advanced shell-model calculations
of \nme~considering the light-neutrino-exchange channel for nuclei of
experimental interest.
We report the sensitivity of the theoretical calculations with
respect to variations in the model spaces and the shell-model nuclear
Hamiltonians.
}


\begin{document}
\section{Introduction}\label{intro}

Neutrinoless double beta (\zbb) decay  is a process in which two
neutrons inside the nucleus transform into two protons with the
emission of two electrons and no neutrinos.
With only two electrons in the final state, lepton number conservation
needs to be violated by two units for the process to occur.
The observation of \zbb\,decay would imply that neutrinos are Majorana
particles---that is, they are their own
antiparticles~\cite{Schechter:1981bd}---and give insight into
leptogenesis scenarios for the generation of the observed
matter-antimatter asymmetry in the universe~\cite{Davidson:2008bu}.
In fact, \zbb\,decay is the most promising laboratory probe of lepton
number violation and it is, in fact, the subject of intense experimental
activities~\cite{Gando:2012zm,Agostini:2013mzu,Albert:2014awa,Andringa:2015tza,KamLAND-Zen:2016pfg,Elliott:2016ble,Agostini:2017iyd,Aalseth:2017btx,Albert:2017owj,Alduino:2017ehq,Agostini:2018tnm,Azzolini:2018dyb}.
The current reported \zbb~half-lives of $^{76}$Ge, $^{130}$Te and
$^{136}$Xe are larger than $8\cdot10^{25}$ yr~\cite{Agostini:2018tnm},
$1.5\cdot10^{25}$ yr~\cite{Azzolini:2018dyb} and $1.1\cdot10^{26}$
yr~\cite{KamLAND-Zen:2016pfg}, respectively, with next generation ton
scale experiments expecting a two orders of magnitude improvement in
the half life sensitivity.

Since \zbb~decay measurements use atomic nuclei as a laboratory to
test the extent of the Standard Model, nuclear theory plays a crucial
role in correctly interpreting the experimental data and disentangling
nuclear physics effects from unknown lepton number violating
mechanisms.
The half-life of the \zbb~decays is given by 
  \begin{equation}
\left[ T^{0\nu}_{1/2}\right]^{-1} = G^{0\nu} \left| M^{0\nu} \right|^2
\left| f(m_i,U_{ei})\right|^2~,
\label{halflife}
\end{equation}
\noindent
where $G^{0\nu}$ is a phase-space (or kinematic) factor~\cite{Kotila12,Kotila13},
\nme~is the nuclear matrix element (NME), and $f(m_i,U_{ei})$  is a function of
the neutrino masses $m_i$ and  their mixing matrix elements $U_{ei}$ that
accounts for beyond-standard-model physics.
Within the light-neutrino exchange mechanisms, $f(m_i,U_{ei})$ has the
following expression:
\[
f(m_i,U_{ei}) = g_A^2 \frac{ \langle m_{\nu}\rangle}{m_e}
\]
\noindent
where $g_A$ is the axial coupling constant, $m_e$ is the electron
mass, and $\langle m _{\nu} \rangle = \sum_i (U_{ei})^2 m_i$ is the
effective neutrino mass.
It is then clear that access to unknown neutrino properties is granted
only if \nme\,  is calculated with great accuracy. 

Currently the calculated matrix elements for nuclei of experimental
interest are characterized by large uncertainties.
For nuclear systems with medium masses and beyond, the many-body
nuclear problem cannot be solved exactly with the available
computational resources.
For these systems, one is inevitably forced to truncate the model
spaces and reduce or neglect the effects of many-body correlations and
electroweak currents. As a result, different computational methods
provide calculated \nmes which differ by a factor of
two~\cite{Engel17}.
On the above grounds, it is clear that reliable calculations of \nme's
are a prime goal of nuclear many-body investigations.

The {\it ab initio} framework for nuclei allows to retain the
complexity of many-body correlations and currents.
Within this approach nuclei are described as systems made of nucleons
interacting via two- and three-body forces. Interactions with external
probes, such as electrons, neutrinos, and photons are described using
many-body current operators.
One- and two-body current operators describe external probes
interacting with individual nucleons and pairs of correlated nucleons,
respectively.
This scheme has been implemented successfully to study light to
intermediate mass nuclei within several many-body computational
approaches.
Due to their prohibitive computational cost, {\it ab initio} methods
have been used to study \zbb\, transitions in light nuclei instead.
While transitions in light nuclei do not have a direct experimental
interest, these studies provide us with an important benchmark to test
other many-body methods that can be used to calculate transition
matrix elements for heavy-mass nuclei of experimental interest.
Further, they allow us to size the importance of the different lepton
number violating mechanisms leading to \zbb\,decay processes, and to
quantify the effect of the various approximations used in the
many-body methods for medium to large nuclear systems.
Studies along this line have been carried out, for example, in
Refs.~\cite{Pastore18,Wang:2019hjy,Basili:2019gvn}.
Only very recently, the {\it ab initio} community is venturing
calculations of \zbb\,decay matrix element of experimental relevance,
as reported, for example, in Ref.~\cite{Hergert16}, where the authors
calculate the  \nme\, of $^{48}$Ca---the lightest system where the
$Q$-value is compatible with the decay---combining the in-medium
similarity renormalization group with the generator-coordinate
method~\cite{Griffin57}.

Besides the exceptions mentioned above, the nuclear physics community
has been primarily focused on employing approximated many-body methods
to access heavy open-shell nuclei of experimental interest.
These approximated methods  generally invoke a truncation of the full Hilbert
space of configurations.
To account for missing dynamics and degrees of freedom in the nuclear
wave functions, the nuclear Hamiltonian is then replaced by an
effective or renormalized Hamiltonian, {\it i.e.}, \heff.
This procedure is carried out, in general, by fitting parameters
inherent the given nuclear model to spectroscopic properties of the
nuclei under investigation.
Nuclear models adopted to study \zbb~decay of nuclei of experimental
interest are: the interacting boson
model~\cite{Barea09,Barea12,Barea13}, the quasiparticle random-phase
approximation (QRPA)~\cite{Simkovic09,Fang11,Faessler12}, the energy
density functional methods~\cite{Rodriguez10}, the covariant
density-functional
theory~\cite{Song14,Yao15,Song17,Jiao17,Yao18,Jiao18,Jiao19}, and the
shell model
(SM)~\cite{Menendez09a,Menendez09b,Horoi13b,Neacsu15,Brown15}.
These models agree within a factor of two (see, for example, Fig. 5 of
Ref. \cite{Engel17} and references therein) when calculating
\zbb~decay matrix elements of $A=48-150$ nuclei.
The difference is mostly to be ascribed to the different
renormalization procedures adopted by the different models.

In addition to renormalize the nuclear Hamiltonian, in this scheme one
has to renormalize  the free constants that appear in the definitions
of the decay operators---{\it e.g.}, proton and neutron electric
charges, spin and orbital gyromagnetic factors, etc.
For example, the axial coupling constant
$g_A^{free}=1.2723$~\cite{PDG18} needs to be quenched by a factor of
$q$~\cite{Towner87}, because all the aforementioned models usually
overestimate Gamow-Teller (GT) rates when compared to the experimental
data~\cite{Chou:1993zz}.
The choice of $q$ depends on the nuclear structure model, the
dimensions of the reduced Hilbert space, and the mass of the nuclei
under investigation~\cite{Suhonen17b}.
The common procedure to handle the quenching of $g_A$ is to fit GT
related data ({\it e.g.}, single-$\beta$ decay strengths, two-neutrino
double-$\beta$ decay rates, etc.), and some authors argue that the
value of $q$ required to reach agreement between theoretical and
experimental values should be also employed to calculate \nme~(see for
instance Refs.~\cite{Suhonen17a,Suhonen17b}).
In passing, it is worth mentioning that  within the {\it ab initio}
framework one can utilize the free nucleonic charges, magnetic
moments, and axial coupling constant without having to resort to
quenching, provided that corrections from two-body currents and
two-body correlations are accounted
for~\cite{Park93,Park:1995pn,Baroni16b,Krebs:2016rqz,Krebs:2020rms,Pastore:2017uwc,King:2020wmp,Gysbers19}.
In this work, we review the most recent and advanced SM results of
\nme~for nuclei currently candidates for the detection of the
\zbb~decay in many laboratories around the world.
We focus on the sensitivity of the calculations with respect to
variations in the model spaces and the shell-model nuclear
Hamiltonians, as well as to the variations in the ``short-range
correlations'' which reveal the role of SM correlated wave functions.

The paper is organized as follows: in Section~\ref{theory} we outline
the basics theory of the nuclear SM and short-range correlations,  and
provide the analytical expressions of the nuclear matrix elements, 
for both neutrinoless and two-neutrino double beta decay.
The latter are reported to assess the validity of the adopted 
nuclear wave functions. In fact, a comparison with experimental data
is clearly possible for two-neutrino double beta decays. 
Section~\ref{calculations} is devoted to the results of the latest SM
calculations for $^{48}$Ca$\rightarrow^{48}$Ti,
$^{76}$Ge$\rightarrow^{76}$Se, $^{82}$Se$\rightarrow^{82}$Kr,
$^{130}$Te$\rightarrow^{130}$Xe, and $^{136}$Xe$\rightarrow^{136}$Ba
\zbb~and \dbb~decays.
Comparisons between experimental and calculated \nmeds~are
reported at the end of Section~\ref{rsm} with a discussion on the $g_A$ quenching.
Our conclusions are given in Section~\ref{conclusions}. 

\section{Theoretical overview}\label{theory}
\subsection{The nuclear shell model}\label{SM}

The nuclear shell model allows for a microscopic description of the
structure of the nucleus~\cite{Mayer49,Haxel49}, and it is the root of
most current {\it ab initio} approaches (No-Core Shell Model, Coupled
Cluster Method, In-Medium Similarity Renormalization Group).
It is based on the ansatz that each nucleon inside the nucleus moves
independently in a spherically symmetric mean field generated by all
other constituents.
The mean field is usually described by a Woods-Saxon or a harmonic
oscillator (HO) potential supplemented by a strong spin-orbit term.

This basic version of the shell model successfully explains the
appearance of protons and/or neutrons ``magic
numbers''---characterizing nuclei bounded more tightly with respect to
their neighbors---along with several nuclear
properties~\cite{Mayer55}, including angular momenta and parity for
ground-states of odd-mass nuclei.
Within this framework, nucleons arrange themselves in well defined and
separated energy levels, {\it i.e.}, the ``shells''.
It is worth emphasizing that shell-model wave functions do not include
correlations induced by the strong short-range two-nucleon interaction.
We will come back to this point later, when we discuss the
``short-range correlations'' (SRC).

The SM can be further improved, especially its description of
low-energy nuclear structure, introducing the ``interacting shell
model'' (ISM) picture.
In the ISM, the complex nuclear many-body problem is reduced to a
simplified one where only few valence nucleons interact in the reduced
model space spanned by a single major shell above an inert core.
The valence nucleons interact via a two-body ``residual interaction'',
that is the part of the interaction which is not already accounted for
in the central potential.
The inclusion of the residual interaction removes the degeneracy of
the states belonging to the same configuration and produces a mixing
of different configurations.

The SM Hamiltonian consists of one- and two-body components and
associated parameters, namely the single-particle (SP) energies and
the two-body matrix elements (TBMEs) of the residual interaction.
These parameters account for the degrees of freedom that are not
explicitly included in the truncated Hilbert space of configurations.
As a matter of fact, SP energies and TBMEs should be determined to
include, in an effective way, the excitations of both the core and the
valence nucleons into the shells above the model space.

The construction of the effective SM Hamiltonian, \heff, can be
carried out into two distinct ways. In one approach one starts from
realistic two- and three-nucleon forces (see
Ref.~\cite{Machleidt:2017vls} and references therein for a review on
realistic two- and three-nucleon potentials)
and derive the effective Hamiltonian from them.
The  \heff~ will then have eigenvalues that belong to the set of
eigenvalues of the full nuclear Hamiltonian, defined in the whole
Hilbert space.
The alternative approach is phenomenological. In this case,  the SM
Hamiltonian one- and two-body components are adjusted to reproduce a
selected set of experimental data.
For the fitting procedure one could i) use adjustable parameters entering the analytical expressions of the residual interaction, ii) directly consider the Hamiltonian matrix 
elements as free parameters (see, {\it
  e.g.},~Refs.~\cite{Elliott69a,Talmi03}), or iii) fine tune the TBMEs
of a realistic \heff~ to reproduce the experimental results.
The phenomenological approach has been widely  utilized since its
formulation in the fifties, and it successfully reproduces a huge
amount of data and describes some of the most fundamental properties
of the structure of atomic nuclei~\cite{Caurier05}.

The SM provides suitable and well tested nuclear wave functions for
the initial and final states entering the calculation of
\nme~associated with \zbb~decays.
SM results based on both the realistic and phenomenological \heff~are
reported in Section~\ref{calculations}.

\subsection{Short-range correlations}\label{SRC}
Short-range correlations (SRCs) are required to account for physics
that is missing in all models that expand nuclear wave functions in
terms of a truncated non-correlated SP basis.
In particular, two-body decay operators---such as those entering
\zbb~decays---acting on an unperturbed (uncorrelated) wave function
yield results that are intrinsically different from those obtained
acting the real (correlated) nuclear wave function~\cite{Bethe71,Kortelainen07}.
Due to the highly repulsive nature of the short-range two-nucleon
interaction, and in order to carry out nuclear structure calculations,
one is forced to perform a consistent regularization of the
two-nucleon potential, $V^{NN}$,  and of any two-body transition
operator~\cite{Wu85}.

\begin{figure}[H]
\begin{center}
\includegraphics[scale=0.7,angle=0]{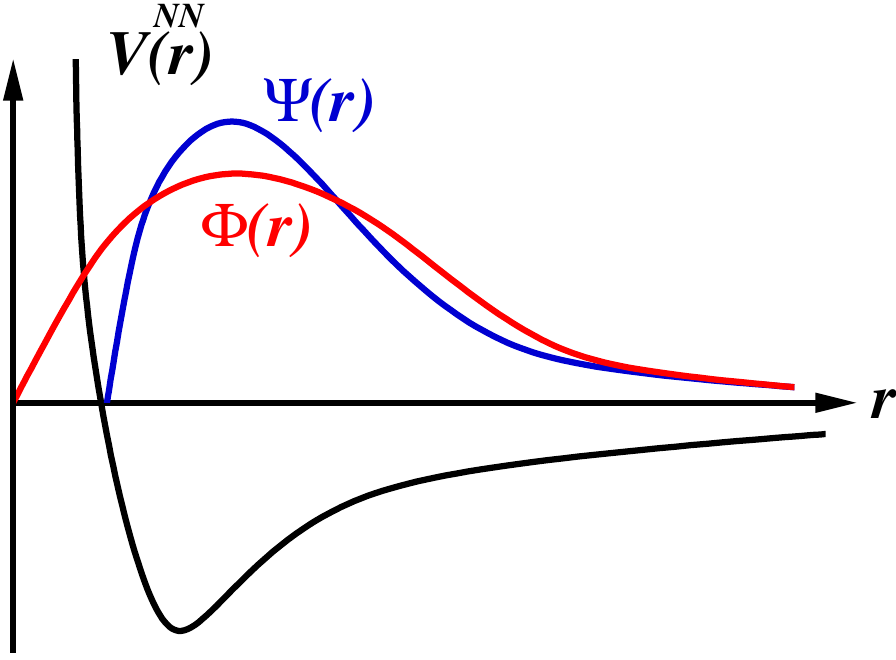}
\end{center}
\caption{Representation of a realistic potential $V^{NN}$, as a
  function of the inter-nucleon distance $r$ (black line). The
  correlated, $\Psi$, and non-correlated, $\Phi$, wave functions are
  represented by the blue and red lines, respectively. See text for
  details.}
\label{psi}
\end{figure}

In nuclear structure calculations based on realistic potentials, one
has to deal with non-zero values of the non-correlated wave function,
$\Phi$, at short distances. This can be appreciated in
Fig.~\ref{psi}.
However, because of the repulsive nature of the $V^{NN}$ interaction
at small inter-particle distances (or equivalently, the repulsive
$V^{NN}$ behavior at high-momentum) the correlated wave function,
$\Psi$, has to approach zero as the inter-nucleon distance diminishes,
and as fast as the core repulsion increases, see Fig.~\ref{psi}. 

To remedy to this shortcoming, one has to renormalize the short-range
(high-momentum) components of the $V^{NN}$ potential whenever a
perturbative approach to the many-body problem is pursued.
The most common way to soften the matrix elements of the
$0\nu\beta\beta$ decay operator is to include SRC given by Jastrow
type functions~\cite{Miller76,Neacsu12}.
Recently, SRC have been modeled using the Unitary Correlation Operator
Method (UCOM)~\cite{Kortelainen07,Menendez09b}.
This approach prevents the overlap between wave functions of a pair of
nucleons~\cite{Feldmeier98}.

Another approach has been proposed by some of the present authors in
Refs.~\cite{Coraggio19b,Coraggio20a}, where the renormalization of the
the \zbb~two-body decay operator is carried out consistently with the
\vlwk~procedure~\cite{Bogner02} adopted to renormalize the repulsive
high-momentum components of the $V^{NN}$ potential.
In particular, the renormalization of $V^{NN}$ occurs through a
unitary transformation, $\Omega$, which decouples the full momentum
space of the two-nucleon Hamiltonian, $H^{NN}$, into two subspaces;
the first one is associated with the relative-momentum configurations
below a cutoff $\Lambda$ and is specified by a projector operator $P$,
the second one is defined in terms of its complement
$Q=\mathbf{1}-P$~\cite{Coraggio19b}.
Being a unitary transformation, $\Omega$ preserves the physics of the
original potential for the two-nucleon system, {\it e.g.}, the
calculated values of all nucleon-nucleon observables are the same as
those reproduced by solving the Schr\"odinger equation for two
nucleons interacting via  $V^{NN}$.

The two-body \zbb~ operator, $\Theta$, is calculated in the momentum
space and renormalized using $\Omega$. This ensure a consistency with
the $V^{NN}$ potential, whose  high-momentum (short range) components
are dumped by the introduction of the cutoff $\Lambda$.
The $\Theta$ vertices appearing in the perturbative expansion of the
$\hat{\Theta}$ box are substituted with the renormalized $\Theta_\lwk$
operator. The latter is defined as $\Theta_\lwk \equiv P \Omega
\Theta \Omega^{-1} P$ for relative momenta $k < \Lambda$, and is set
to zero for $k > \Lambda$.
The magnitude of the overall effect of this renormalization procedure
is comparable to using the SRC modeled by the Unitary Correlation
Operator Method~\cite{Menendez09b}, that is a lighter softening of
\nme~with respect to the one provided by Jastrow type
SRC~\cite{Coraggio19b}.

\subsection{The \zbb-decay operator for the light-neutrino
  exchange}\label{0vv}
We now turn our attention to the vertices of the bare \zbb~operator,
$\Theta$, for the light-neutrino-exchange channel~\cite{Engel17}.

We recall that the formal expression of $M_{\alpha}^{0\nu}$---where
$\alpha$ stands for Fermi ($F$), Gamow-Teller (GT), or tensor ($T$)
decay channels---is written in terms of the one-body
transition-density matrix elements between the daughter and parent
nuclei (grand-daughter and daughter nuclei) $ \langle k |
a^{\dagger}_{p^\prime} a_{n^\prime} | i \rangle$ ($ \langle f |
a^{\dagger}_{p}a_{n} | k \rangle $).
Here, $p$ and $n$ denote proton and neutron states, and $i,k,f$ refer
to the parent, daughter, and grand-daughter nuclei, respectively,
while $M_{\alpha}^{0\nu}$ reads \cite{Senkov13,Simkovic08}:
\begin{align}
  M_\alpha^{0\nu} =  
&\sum_{k {\cal J}} \sum_{j_p j_{p^\prime} j_n
    j_{n^\prime} } (-1)^{j_n + j_{n^\prime}+ \cal{J}} \hat{\cal{J}} \left\{
\begin{array}{ccc}
j_p & j_n & J_\kappa \\
j_{n^\prime} & j_{p^\prime} & {\cal J}
\end{array}
\right\} \left< j_p  j_{p^\prime} ;{\cal J} \mid \mid \tau^-_{1} \tau^-_{2}
\Theta_\alpha^{k} \mid \mid j_n
 j_{n^\prime}; {\cal J} \right>  \langle k || [a^{\dagger}_{p} \otimes
                              \tilde{a}_{n}]_{J_k} || i \rangle
                              \nonumber \\
~& \langle k || [a^{\dagger}_{n^\prime} \otimes \tilde{a}_{p^\prime}]_{J_k} || f
   \rangle ^{\ast} =
\sum_{k} \sum_{j_p j_{p^\prime} j_n  j_{n^\prime} }
   \langle f | a^{\dagger}_{p}a_{n} | k \rangle \langle k |
   a^{\dagger}_{p^\prime} a_{n^\prime} | i \rangle \times \left< j_p
   j_{p^\prime} \mid \tau^-_{1}
\tau^-_{2} \Theta_\alpha^{k} \mid j_n j_{n^\prime} \right> ~, \label{M0nu}
\end{align}

where the tilde denotes a time-conjugated state, $\tilde{a}_{j m} =
(-1)^{j+m}a_{j -m}$.

The operators $\Theta_{\alpha}$ given by~\cite{Engel17}:
\begin{eqnarray} 
 \Theta^{k}_{\rm GT} & = & \vec{\sigma}_1 \cdot \vec{\sigma}_2 H_{\rm
GT}^k(r) \label{operatorGT} \, , \\
\Theta^{k}_{\rm F} & = & H_{\rm F}^k(r)  \label{operatorF} \, ,\\ 
\Theta^{k}_{\rm T} & = & \left[3\left(\vec{\sigma}_1 \cdot \hat{r} \right) 
\left(\vec{\sigma}_1 \cdot \hat{r} \right) - \vec{\sigma}_1 \cdot
                      \vec{\sigma}_2 \right] H_{\rm T}^k(r)~, \label{operatorT}
\end{eqnarray}
where $H_{\alpha}$ are the  neutrino potentials defined as:
\begin{equation}
H_{\alpha}^k(r)=\frac {2R}{\pi} \int_{0}^{\infty} \frac {j_{n_{\alpha}}(qr)
  h_{\alpha}(q^2)qdq}{q+E_k-(E_i+E_f)/2}~.
\label{neutpot}
\end{equation}
In the equation above,  $R=1.2 A^{1/3}$ fm, $j_{n_{\alpha}}(qr)$ is
the spherical Bessel function, $n_{\alpha}=0$ for Fermi and
Gamow-Teller components, while $n_{\alpha}=2$ for the tensor
component.
For the sake of clarity, the explicit expressions~\cite{Engel17} of
neutrino form functions, $h_{\alpha}(q)$, for light-neutrino exchange
are reported below:
\begin{eqnarray}
h_{\rm F} ({ q}^{2})  & = & g^2_V({ q}^{2}) \, ,  \nonumber \\
h_{\rm GT} ({ q}^{2}) & = & \frac{g^2_A({ q}^{2})}{g^2_A} 
\left[ 1 - \frac{2}{3} \frac{ { q}^{2}}{ { q}^{2} + m^2_\pi } + 
\frac{1}{3} ( \frac{ { q}^{2}}{ { q}^{2} + m^2_\pi } )^2 \right]
\nonumber\\
&& + \frac{2}{3} \frac{g^2_M({ q}^{2} )}{g^2_A} \frac{{ q}^{2} }{4 m^2_p }, 
\nonumber \\
h_{\rm T} ({ q}^{2}) & = & \frac{g^2_A({ q}^{2})}{g^2_A} \left[ 
\frac{2}{3} \frac{ { q}^{2}}{ { q}^{2} + m^2_\pi } -
\frac{1}{3} ( \frac{ { q}^{2}}{ { q}^{2} + m^2_\pi } )^2 \right] 
\nonumber\\
&& + \frac{1}{3} \frac{g^2_M ({ q}^{2} )}{g^2_A} \frac{{ q}^{2} }{4 m^2_p }  \, ,   
\end{eqnarray}
For the vector, $g_V({ q}^{2})$, axial-vector, $g_A({ q}^{2})$, and
weak-magnetism, $g_M({ q}^{2})$, form factors we use the dipole
approximation:
\begin{eqnarray} 
g_V({ q}^{2})&=& \frac{g_V}{(1+{ q}^{2}/{\Lambda^2_V})^2}, 
\nonumber\\
g_M({ q}^{2}) &=& (\mu_p-\mu_n) g_V({ q}^{2}), 
\nonumber\\
g_A({ q}^{2}) &=& \frac{g_A}{(1+{ q}^{2}/{\Lambda^2_A})^2},
\end{eqnarray}
where $g_V = 1$, $g_A \equiv g_A^{free}=1.2723$,  $(\mu_p - \mu_n) =
4.7$, and the cutoff parameters $\Lambda_V = 850$ MeV and $\Lambda_A =
1086$ MeV.

The total nuclear matrix element \nme~can be then written as
\begin{equation}
M^{0\nu} =  M_{\rm GT}^{0\nu} - \frac{g_V^2}{g_A^2}  M_{\rm F}^{0\nu}
+  M_{\rm T}^{0\nu}~~.
\end{equation}

The expression in Eq.~(\ref{M0nu}) can be easily calculated within the
QRPA computational approach, while all other models---including most
of the SMs---have to resort to the closure approximation.
This approximation is based on the observation that the relative
momentum $q$ of the neutrino, appearing in the propagator of
Eq.~(\ref{neutpot}), is of the order of 100-200 MeV~\cite{Engel17},
while the excitation energies of the nuclei involved in the transition
are of the order of 10 MeV~\cite{Senkov13}. It is then customary to
replace the energies of the intermediate states, $E_k$, appearing in
Eq.~(\ref{neutpot}), by an average value $E_k-(E_i+E_f)/2 \rightarrow
\langle E \rangle$. This allow us to simplify both Eqs.~(\ref{M0nu})
and~(\ref{neutpot}).
In particular, $M_\alpha^{0\nu}$ can be re-written in terms of the
two-body transition-density matrix elements $\langle f |
a^{\dagger}_{p}a_{n} a^{\dagger}_{p^\prime} a_{n^\prime} | i \rangle$
as
\begin{eqnarray}
M_\alpha^{0\nu}& =  & \sum_{j_n j_{n^\prime} j_p j_{p^\prime}}
\langle f | a^{\dagger}_{p}a_{n} a^{\dagger}_{p^\prime} a_{n^\prime} 
| i \rangle \nonumber \\
~ & ~& \times  \left< j_p  j_{p^\prime} \mid \tau^-_{1}
\tau^-_{2} \Theta_\alpha \mid  j_n j_{n^\prime}
       \right>~, \label{M0nuapp}
\end{eqnarray}
and the neutrino potentials become
\begin{equation}
H_{\alpha}(r)=\frac {2R}{\pi} \int_{0}^{\infty} \frac {j_{n_{\alpha}}(qr)
  h_{\alpha}(q^2)qdq}{q+\left< E \right>}~.
\label{neutpotapp}
\end{equation}
Most SM calculations adopt the closure approximation to define the
$\Theta$ operators given in
Eqs.~(\ref{operatorGT})--(\ref{operatorT}), and take the average
energies $\left< E \right>$ from the evaluations of
Refs.~\cite{Haxton84,Tomoda91}. 
It is important to point out that the authors of Ref.~\cite{Senkov13}
performed SM calculations of \nme~for $^{48}$Ca both within and beyond
the closure approximation, and found that in the second case the
results are $\sim 10\%$ larger.

In most cases, short-range correlations are included when computing
the radial matrix elements of the neutrino potentials $\left <
  \psi_{nl}(r) |H_{\alpha}| \psi_{n^\prime l^\prime}(r) \right>$.
In particular, the HO wave functions  $\psi_{nl}(r)$ and
$\psi_{n^\prime l^\prime}(r)$ are corrected by a factor $[1 + f(r)]$,
which takes into account the short range correlations induced by the
nuclear interaction
\begin{equation}
\psi_{nl}(r) \rightarrow \left[ 1+f(r) \right] \psi_{nl}(r) \ .
\end{equation}
The functional form of the correlation function $f(r)$ is usually
written using a Jastrow-like parametrization as~\cite{Neacsu12}
\begin{equation}
f(r) = - c \cdot e^{-ar^2} \left( 1-br^2 \right) ~~,
\label{functionalSRC}
\end{equation}
where $a$, $b$ and $c$ are parameters whose values depend on the
renormalization procedure adopted to renormalize the non-correlated HO
wave functions, ({\it e.g.}, Jastrow or UCOM schemes, see
Section~\ref{SRC} for details).
In Table~\ref{SRCvalues} we report the values of the $a$, $b$ and $c$
constants commonly employed in SM calculations.
In addition to the values proposed by Miller and
Spencer~\cite{Miller76}, we show those based on the modern
nucleon-nucleon interactions CD-Bonn and AV18 and derived in
Ref.~\cite{Simkovic09}.
\begin{table}[H]
\caption{Values of the SRC parameters.}\label{SRCvalues}
\centering
\begin{tabular}{lccc}
\toprule
 ~& \textbf{a}	& \textbf{b}	& \textbf{c}\\
\midrule
Miller-Spencer & 1.10		& 0.68			& 1.00\\
CD-Bonn & 1.52		& 1.88			& 0.46\\
AV18 & 1.59		& 1.45			& 0.92\\
\bottomrule
\end{tabular}
\end{table}

\subsection{The \dbb-decay operator}\label{2vv}
As already pointed out in the Introduction, because of the
impossibility to compare the theoretical values of \nme~ with the
experiment, one has to find another way to check the reliability of
the computed results.
A viable route that is often considered in literature is the
calculation of the standard or ordinary two-neutrinos double beta
decay transitions where one observes the emission of two electrons and
two antineutrinos.
Two-neutrino double beta decays are simply the occurrence of two
single beta decay transitions inside a nucleus.
They differ from \zbb\, decays in the characteristic value of momentum
transfer, which is  negligible in ordinary decays and of the order of
hundreds of MeVs in \zbb\, decay.
Here, we list the expressions of the GT and Fermi components of the 
two-neutrinos double beta decay matrix elements $M^{2\nu}$, namely
\begin{eqnarray}
M_{\rm GT}^{2\nu} & = & \sum_n \frac{ \langle 0^+_f || \vec{\sigma} \tau^-
  || 1^+_n \rangle \langle 1^+_n || \vec{\sigma}
\tau^- || 0^+_i \rangle } {E_n + E_0} ~~,\label{doublebetameGT} \\
M_{\rm F}^{2\nu} & = & \sum_n \frac{ \langle 0^+_f || \tau^-
  || 0^+_n \rangle \langle 0^+_n || \tau^- || 0^+_i \rangle } {E_n +
  E_0} ~~.\label{doublebetameF}
\end{eqnarray}

\noindent
In the equation above, $E_n$ is the excitation energy of the $J^{\pi}=0^+_n,1^+_n$ 
intermediate state, and $E_0=\frac{1}{2}Q_{\beta\beta}(0^+) +\Delta M$, where
$Q_{\beta\beta}(0^+)$ and $\Delta M$ are the $Q$ value of the transition
and the mass difference of the initial and final nuclear states, respectively.
The index $n$ runs over all possible intermediate states induced by the given 
transition operator.

It should be pointed out that the Fermi component is zero in Hamiltonians that 
conserve the isospin, and most of the SM effective Hamiltonians do. It would 
play a marginal role only when isospin violation mechanisms are introduced, 
for example, to account for the effects of the Coulomb force acting between the
valence protons \cite{Haxton84,Elliott02}. In practice, in most calculations, the
Fermi component is neglected altogether. 

An efficient way to calculate \nmed~is to resort to the Lanczos
strength-function method \cite{Caurier05}, which allows to include the 
intermediate states required to obtain a given accuracy for the
calculated values.

The theoretical values are then compared with the experimental
counterparts, that are extracted from the observed
half life $T^{2\nu}_{1/2}$ 

\begin{equation}
\left[ T^{2\nu}_{1/2} \right]^{-1} = G^{2\nu} \left| M_{\rm GT}^{2\nu}
\right|^2 ~~.
\label{2nihalflife}
\end{equation}

One can base the evaluation of the $M_{\rm GT}^{2\nu}$  on the 
closure approximation, commonly adopted to study \zbb-decay NMEs 
\cite{Haxton84}. Within this approximation, one can avoid to 
explicitly calculate the intermediate $J^{\pi}=1^+_n$ states.
The drawback is that, in using the closure on the intermediate states,
the two one-body transition operators become a two-body operator.  

This approximation is more adapt to evaluate \nme where the 
neutrino's momentum is about one order of magnitude greater than the average 
excitation energy of the intermediate states. This allows to safely 
neglect intermediate-state-dependent energies from the energy 
denominator appearing in the neutrino potential (see discussion in 
Section \ref{0vv}). Conversely, the closure approximation has turned 
out to be unsatisfactory when used to calculate  \nmed, and that is 
because the momentum transfer in \dbb~process are much smaller. Theoretical
calculations of \nmed are discussed in the next session.  \\

\section{Shell-model results}\label{calculations}
In this Section, we report SM results for \nme~ based on both the
phenomenological and realistic \heff's.
All the calculations are based on the light-neutrino-exchange
hypothesis, and the values of all the input parameters are the same as
reported in Section \ref{0vv}.
The only exception is the $g_A$ parameter, whose adopted value is
equal to 1.254 in some reported calculations.
It is worth pointing out, however, that in Ref. \cite{Neacsu13}, where
it can be found a detailed analysis of the sensitivity of the \nme
results on the values of the input parameters, it has been shown that
the effects of such a tiny difference in $g_A$ are negligible.
We will focus our attention on the $^{48}$Ca, $^{76}$Ge, $^{82}$Se,
$^{128}$Te, $^{130}$Te and $^{136}$Xe emitters.
These results have been obtained performing a complete
diagonalizations of \heff. The latter has been defined in different
valence spaces tailored for the specific decay under investigation.
All the calculations based on phenomenological interactions are
performed starting from Brueckner $G$-matrix elements ``fine tuned''
to reproduce some specific set of spectroscopic data.

\subsection{Results from phenomenological \heff's}
\label{empirical}
We test different phenomenological \heff's.
All these interactions have been derived modifying the matrix elements
of a $G$-matrix so as to reproduce a chosen set of spectroscopic
properties of some nuclei belonging to the mass region of interest.
With this procedure one can end up  with results that provide similar
descriptions of the nuclei under consideration, nevertheless the
phenomenological TBMEs are quite different each other.

It is worth stressing out that the calculated \nmes~, reported in this
section, are obtained using free value of the axial coupling constant
$g_A$ without any quenching factor.

The double-magic nucleus $^{48}$Ca is the lightest emitter
investigated in regular $\beta\beta$ decay searches.
The SM calculation for \nme~is obtained using the model space spanned
by four neutron and proton single-particle orbitals 0f$_{7/2}$,
1p$_{3/2}$, 1p$_{1/2}$, and 0f$_{5/2}$.
It is worth mentioning that the regular $\beta\beta$ decay of
$^{48}$Ca is a paradigm for shell-model calculations. Because within
the $pf$ model space all spin-orbit partners are present, the Ikeda
sum rule is satisfied~\cite{Ikeda64}.

Several phenomenological SM effective interactions have been developed
to describe $pf$-shell nuclei.
Among these are the GXPF1~\cite{Honma04}, GXPF1A~\cite{Honma05},
KB3~\cite{Poves81}, KB3G~\cite{Poves01}, and FPD6~\cite{Richter91}
interactions.
In Table~\ref{48Canme}, we compare the most recent results for the
\nme~ of $^{48}$Ca obtained using the GXPF1A \cite{Senkov13} and KB3G
interactions \cite{Menendez18}.
\begin{table}[H]
\caption{\nme~of $^{48}$Ca. (a) and
  (b) denote AV18 and CD-Bonn SRC parametrizations,
  respectively. The results have been taken from
  Refs.~\cite{Senkov13,Menendez18}}\label{48Canme} 
\centering
\begin{tabular}{lccc}
\toprule
 ~& GXPF1A (a)	& KB3G (a)	& KB3G (b) \\
\midrule
$M_{\rm GT}^{0\nu}$ & 0.68		& 0.85			& 0.93 \\
$M_{\rm F}^{0\nu}$  & -0.20		& -0.23			& -0.25\\
$M_{\rm T}^{0\nu}$  & -0.08		& -0.06			& -0.06\\
$M^{0\nu}$     & 0.73		& 0.93			& 1.02\\
\bottomrule
\end{tabular}
\end{table}
For the medium-mass emitters $^{76}$Ge and $^{82}$Se, the calculations
adopt the valence space with the four neutron and proton single-particle
orbitals 0f$_{5/2}$, 1p$_{3/2}$, 1p$_{1/2}$, 0g$_{9/2}$
outside doubly-magic $^{56}$Ni, as for instance in
Refs. \cite{Senkov14} and \cite{Menendez18}, where the effective
interactions GCN2850 \cite{Menendez09b} and JUN45 \cite{Honma09} have
been employed.
These results are given in Tables \ref{76Genme} and~\ref{82Senme}.

\begin{table}[H]
  \caption{Same as Table~\ref{48Canme}, but for $^{76}$Ge.
    Results are from
    Refs.~\cite{Senkov14,Menendez18}.}\label{76Genme}
\centering
\begin{tabular}{lcccc}
  \toprule
  & JUN45 (a)   & JUN45 (b) & GCN2850 (a) & GCN2850 (b) \\ 
\midrule
$M_{\rm GT}^{0\nu}$ & 2.98   & 3.15		& 2.56			& 2.73 \\
$M_{\rm F}^{0\nu}$  & -0.62  & -0.67		& -0.54			& -0.59\\
$M_{\rm T}^{0\nu}$  & -0.01  & -0.01		& -0.01			& -0.01\\
$M^{0\nu}$     & 3.15   &  3.35		& 2.89			& 3.07\\
\bottomrule
\end{tabular}
\end{table}

\begin{table}[H]
\caption{Same as Table~\ref{48Canme}, but for $^{82}$Se. Results
  are from Refs. \cite{Senkov14,Menendez18}.}\label{82Senme} 
\centering
\begin{tabular}{lccc}
\toprule
  & JUN45 (b) & GCN2850 (a) & GCN2850 (b) \\ 
\midrule
$M_{\rm GT}^{0\nu}$ & 2.75		& 2.41			& 2.56 \\
$M_{\rm F}^{0\nu}$  & -0.61		& -0.51			& -0.55\\
$M_{\rm T}^{0\nu}$  & -0.01		& -0.01			& -0.01\\
$M^{0\nu}$     &  3.13		& 2.73			& 2.90\\
\bottomrule
\end{tabular}
\end{table}

Finally, in Tables \ref{130Tenme} and \ref{136Xenme} we report and
compare  the calculated \nme~ for $^{130}$Te and $^{136}$Xe.
These are based on two different effective interactions, namely the
SVD~\cite{Qi12} and the GCN5082~\cite{Menendez09b} defined in the
$jj55$ valence space spanned by the neutron and proton orbitals
0g$_{7/2}$, 1d$_{5/2}$, 1d$_{3/2}$, 2s$_{1/2}$, and 0h$_{11/2}$.
Results with the SVD and GCN5082 interactions are taken, respectively,
from Refs. \cite{Neacsu15} and \cite{Menendez18}.

\begin{table}[H]
  \caption{Same as Table \ref{48Canme}, but for $^{130}$Te.
    Results are from
    Refs.~\cite{Neacsu15,Menendez18}.}\label{130Tenme}  
\centering
\begin{tabular}{lcccc}
  \toprule
  & SVD (a)   & SVD (b) & GCN5082 (a) & GCN5082 (b) \\ 
\midrule
$M_{\rm GT}^{0\nu}$ & 1.54   & 1.66		& 2.36			& 2.54 \\
$M_{\rm F}^{0\nu}$  & -0.40  & -0.44		& -0.62			& -0.67\\
$M_{\rm T}^{0\nu}$  & -0.01  & -0.01		& 0.00			& 0.00\\
$M^{0\nu}$     & 1.80   &  1.94		& 2.76			& 2.96\\
\bottomrule
\end{tabular}
\end{table}

\begin{table}[H]
  \caption{Same as Table \ref{48Canme}, but for $^{136}$Xe.
    Results are from
    Refs.~\cite{Neacsu15,Menendez18}.}\label{136Xenme}  
\centering
\begin{tabular}{lcccc}
  \toprule
    & SVD (a)   & SVD (b) & GCN5082 (a) & GCN5082 (b) \\ 
\midrule
$M_{\rm GT}^{0\nu}$ & 1.39   & 1.50		& 2.56			& 2.73 \\
$M_{\rm F}^{0\nu}$  & -0.37  & -0.40		& -0.50			& -0.54\\
$M_{\rm T}^{0\nu}$  & -0.01  & -0.01		& 0.00			& 0.00\\
$M^{0\nu}$     & 1.63   &  1.76		& 2.28			& 2.45\\
\bottomrule
\end{tabular}
\end{table}

From the results shown above, it can be inferred that the effect
associated with using different SRCs does not exceed 10\%, while
different effective interactions can provide results differing up to
50\%.
The results reported in this Section are based on the closure
approximation.
As discussed above, Senkov and coworkers have shown in a series of
papers~\cite{Senkov13,Senkov14,Senkov16} that in going beyond this
approximation, the \zbb~decay~\nme becomes about $10\%$ larger.

We recall that the results reported in Tables
\ref{48Canme}-\ref{136Xenme} are obtained without quenching the axial
coupling constant.
However, the calculations based on the empirical SM Hamiltonians so
far considered need a quenching factor $q$ different from 1 
to reproduce the experimental values of the nuclear matrix elements of
the corresponding \dbb-decays \nmeds.
This can be appreciated in Table \ref{tablenmed} where we list the 
\nmeds~calculated with the empirical effective Hamiltonians GXPF1A,
KB3G, JUN45, GCN2850 and GCN5082, and compare them with the
experimental data.
In these calculations, that are performed employing the Lanczos
strength-function method \cite{Caurier05}, it has been used the
unquenched value of $g_A$ (or equivalently a quenching factor $q=1$),
and, as expected, the theory is systematically overpredicting the
experimental data.

\begin{table}[H]
\caption{\nmeds~for $^{48}$Ca, $^{76}$~Ge, $^{82}$Se, $^{130}$Te, and
  $^{136}$Xe \dbb-decay calculated with GXPF1A, KB3G, JUN45, GCN2850,
  and GCN5082 interactions and compared with data \cite{Barabash20},
  there are no published results based on the SVD interaction.
  These calculations use a quenching factor $q=1$.
  The values of \nmeds~ are reported in MeV$^{-1}$.}\label{tablenmed}
\centering
\begin{tabular}{lccc}
\toprule
 $^{48}$Ca$\rightarrow^{48}$Ti & GXPF1A & KB3G & Expt.\\
~ & 0.0511 \cite{Kostensalo20} & 0.088 \cite{Caurier12} & $0.035\pm 0.003$ \\
\midrule
 $^{76}$Ge$\rightarrow^{76}$Se & JUN45 & GCN2850 & Expt.\\
~ & 0.333 \cite{Caurier12} & 0.322 \cite{Caurier12} & $0.106\pm 0.004$ \\
\midrule
 $^{82}$Se$\rightarrow^{82}$Kr & JUN45 & GCN2850 & Expt.\\
~ & 0.344 \cite{Caurier12} & 0.350 \cite{Caurier12} & $0.085\pm 0.001$ \\
\midrule
 $^{130}$Te$\rightarrow^{130}$Xe & GCN5082 & ~ & Expt. \\
~ & 0.132 \cite{Caurier12} & ~& $0.0293\pm 0.0009$ \\
\midrule
 $^{136}$Xe$\rightarrow^{136}$Ba & GCN5082 & ~ & Expt. \\
~ & 0.123 \cite{Caurier12} & ~ & $0.0181\pm 0.0006$ \\
\bottomrule
\end{tabular}
\end{table}

\subsection{Results from realistic \heff's}\label{rsm}
In the realistic SM (RSM), \heff~ is constructed from realistic
$V^{NN}$ potentials.
This is achieved via a similarity transformation utilized to constrain
both the SM Hamiltonian and the  SM transition operators.
More details on this procedure can be found in the papers by
B. H. Brandow~\cite{Brandow67}, T. T. S. Kuo and
coworkers~\cite{Kuo71,Krenciglowa74}, and K. Suzuki and
S. Y. Lee~\cite{Lee80,Suzuki80}.
Perturbative and non-perturbative derivations of \heff~ have been most
recently reviewed in Ref.~\cite{Coraggio12a} and
Ref.~\cite{Stroberg19}, respectively.
The derivation of effective SM decay operators carried out 
consistently with \heff~is discussed in Refs.~\cite{Ellis77,Suzuki95}.
Fundamental contributions to the field have been made by 
I. S. Towner, who has extensively investigated the role of many-body
correlations induced by the truncation of the Hilbert space,
especially for spin- and spin-isospin-dependent one-body decay
operators \cite{Towner83,Towner87}.

The first calculations of \nme~ starting from realistic $V^{NN}$
potentials and associated effective SM decay operators, were made by Kuo
and coworkers in the middle of 1980's for $^{48}$Ca~\cite{Wu85}.
In that reference, \heff\, and the associated transition operators
were based on the Paris and the Reid potentials~\cite{Lacombe80,Reid68}.
The short-range repulsive behavior was renormalized by calculating the
corresponding Brueckner reaction matrices~\cite{Krenciglowa76}.
Many-body perturbation theory was then implemented to derive
both the TBMEs of \heff~ and the effective \zbb-decay operator.
The effect of the SRC was embedded in the defect wave function
\cite{Kuo66},  consistently with the renormalization procedure from the
Paris and Reid potentials.
Finally, the authors calculated the half lives of $^{48}$Ca \zbb-decay,
for  both  light- and heavy-neutrino exchange, as a function of the
neutrino effective mass.

In more recently, J. D. Holt and J. Engel calculated effective SM
operators  \teff's from modern chiral effective field theory 
$V^{NN}$ potentials. In particular, they started from the chiral
$V^{NN}$ potential by Entem and Machleidt \cite{Entem02} and
cured its perturbative behavior using the \vlwk~procedure~\cite{Bogner02}. 
The \teff~was expanded up to third order in perturbation theory and used 
to calculate \nme~for $^{76}$Ge, $^{82}$Se\cite{Holt13d}, and
$^{48}$Ca\cite{Kwiatkowski14}. The effects of SRC was included via an
effective Jastrow function obtained from Brueckner-theory calculations
\cite{Simkovic09}.
For $^{76}$Ge and $^{82}$Se decays, the authors employed the
empirical GCN2850~\cite{Menendez09b} and JUN45~\cite{Honma09} SM
interactions, respectively, and for the \zbb~decay of $^{48}$Ca  they
used the GXPF1A \heff~\cite{Honma05}.
The values obtained by Holt and Engel in the light-neutrino exchange
channel are \nme=1.30 for $^{48}$Ca , \nme=3.77 for $^{76}$Ge, and
\nme=3.62 for $^{82}$Se~\cite{Holt13d,Kwiatkowski14}.

In Ref. \cite{Holt13d}, the authors calculated also the $^{76}$Ge \dbb~
matrix element. The calculation used the closure approximation.
However, as we discussed in Sessions \ref{0vv} and \ref{2vv},
this approximation is not robust when applied to study  
\dbb~processes where the values of momentum transfer are small.
In fact, the authors obtain a result for $^{76}$Ge $M_{\rm
  GT}^{2\nu}$ that is about two times larger than the one calculated
with the Lanczos strength-function method~\cite{Horoi13c,Brown15},
and about 5 times larger than the experimental value
\cite{Barabash20}.

RSM calculations based on the high-precision CD-Bonn $NN$
potential~\cite{Machleidt01b} have been recently carried out in
Ref.~\cite{Coraggio20a}, where the repulsive high-momentum components
have been integrated out through the \vlwk~technique with ``hard
cutoff'' $\Lambda=2.6$ fm$^{-1}$~\cite{Bogner02}.
The \nme's~have been calculated within the SM using \heff, and
effective decay operators \teff's~up to the third order in
perturbation theory.
Two-body matrix elements entering the \zbb-decay operator have been
renormalized consistently within the \vlwk~to account for short-range
correlations (see Section \ref{SRC} for more details), and
Pauli-principle violations in the effective SM operator.
It should be pointed out that calculations of systems with a number
$n$ of valence nucleons require the derivation of $n$-body effective
operators, that take into account the evolution of the number of
valence particle in the model space $P$.
The correlation between $P$-space configurations and those belonging
to $Q$ space is affected by the filling of the model-space orbitals,
and in a perturbative expansion of SM operators this is considered by
way of $n$-body diagrams.
This is called the ``Pauli-blocking effect'' and calculations in
Ref. \cite{Coraggio20a}, where all SM parameters have been
consistently derived from the realistic $V^{NN}$ potential without any
empirical adjustments, take it into consideration by including the
contribution of three-body  correlation diagrams to derive \teff.

The results for \zbb~decay in the light-neutrino exchange channel of
$^{48}$Ca, $^{76}$Ge, $^{82}$Se, $^{130}$Te, and $^{136}$Xe are
reported in Table \ref{NMEreal}.

\begin{table}[ht]
  \caption{ \nmes~of $^{48}$Ca, $^{76}$Ge, $^{82}$Se, $^{130}$Te, and
    $^{136}$Xe decays obtained with the realistic effective
    Hamiltonians and operators \cite{Coraggio20a}.
    The tensor component has been neglected.}
\label{NMEreal}
\centering
\begin{tabular}{lccc}
\toprule
 Decay & $M^{0\nu}_{\rm GT}$ & $M^{0\nu}_{\rm F}$  & $M^{0\nu}$ \\
\midrule
~ & ~ & ~& ~ \\
  $^{48}$Ca  $\rightarrow$ $^{48}$Ti & 0.22 & -0.12 & 0.30 \\
  \midrule
  $^{76}$Ge  $\rightarrow$ $^{76}$Se & 2.25 & -0.65 & 2.66 \\
  \midrule
  $^{82}$Se  $\rightarrow$ $^{82}$Kr & 2.31 & -0.66 & 2.72 \\
  \midrule
  $^{130}$Te $\rightarrow$ $^{130}$Xe & 2.66 & -0.80 & 3.16 \\
  \midrule
$^{136}$Xe  $\rightarrow$ $^{136}$Ba & 2.01 & -0.61 & 2.39 \\
\bottomrule
\end{tabular}
\end{table}

It is worth pointing out that this approach to SM calculations has
been successfully tested on energy spectra, electromagnetic transition
strengths, GT strength distributions, and nuclear matrix elements for
the two-neutrino $\beta\beta$ decay~\cite{Coraggio17a, Coraggio19a},
without resorting to effective proton/neutron charges and gyromagnetic
factors, or quenching of $g_A$.
In Table \ref{tablenmedreal} we report the the calculated values of
\nmed~ from Ref.~\cite{Coraggio19a}
and compare them with experimental data~\cite{Barabash20}.
\begin{table}[H]
\caption{Same as in Table \ref{tablenmed}, but the calculated values
  \cite{Coraggio19a} are obtained employing effective Hamiltonians and
  decay operators derived starting from the CD-Bonn realistic
  potential (see text for details) and compared with experiment
  \cite{Barabash20}.}\label{tablenmedreal}
\centering
\begin{tabular}{lcc}
\toprule
 $^{48}$Ca$\rightarrow^{48}$Ti & Theory & Expt.\\
~ & 0.026 & $0.035\pm 0.003$ \\
\midrule
 $^{76}$Ge$\rightarrow^{76}$Se & Theory & Expt.\\
~ & 0.104 & $0.106\pm 0.004$ \\
\midrule
 $^{82}$Se$\rightarrow^{82}$Kr & Theory & Expt.\\
~ & 0.109 & $0.085\pm 0.001$ \\
\midrule
 $^{130}$Te$\rightarrow^{130}$Xe & Theory & Expt. \\
~ & 0.061 & $0.0293\pm 0.0009$ \\
\midrule
 $^{136}$Xe$\rightarrow^{136}$Ba & Theory & Expt. \\
~ & 0.0341 & $0.0181\pm 0.0006$ \\
\bottomrule
\end{tabular}
\end{table}

A few comments are now in order.
As we already pointed out in the Introduction, SM calculations
overestimate \nmed~and Gamow-Teller transition strengths.
To remedy to this deficiency one introduces a quenching factor $q$
that is multiplied to $g_A$ to reduce the values of the calculated
matrix elements.
This factor depends on {\it i}) the mass region of the nuclei involved in
the decay process; and {\it ii}) the dimension of the model space used 
in the calculation. The quenching factor has on average the
empirical value $q \approx 0.7$ \cite{Suhonen17b}).

The quenching factor accounts for missing correlations and missing 
many-body effects in the transition operators. The truncation of the
full Hilbert space to the reduced SM space has the effect of excluding
all correlations between the model-space configurations and the
configurations belonging to either the doubly-closed core or the
shells placed in energies above the SM space.
In addition, SM calculations are based on the single-nucleon paradigm
for the transition operators.
However, two-body electroweak currents~\cite{Park93,Park:1995pn,Baroni16b,Krebs:2016rqz,Krebs:2020rms,Pastore:2017uwc,King:2020wmp,Gysbers19,Pastore:2008ui,Pastore09,Pastore:2011ip,Pastore:2014oda,Pastore:2012rp,Piarulli13,Kolling:2009iq,Kolling:2011mt,Bacca:2014tla}
are found to play a role in several electroweak observables.
These involve the exchange of mesons and nucleonic excitations. 

I. S. Towner extensively studied how to construct 
effective $\beta$-decay operators that account for
the degrees of freedom that are not explicitly included in the model
space (see Ref. \cite{Towner87}). This has been more recently investigated in 
Refs. \cite{Coraggio17a, Coraggio19a}).
The results that are reported in Table \ref{tablenmedreal} demonstrate
a satisfactory agreement with the data can be obtained without
resorting to quenching factors $q$ if one employs effective GT
operators within the SM.

Moreover, in Ref. \cite{Coraggio20a} the authors showed that, 
the renormalization procedure implemented to account for missing
configurations plays a marginal role in the calculated \nmes, while it is
relevant in \dbb-decay induced by the one-body GT operator. 
This evidences that the mechanisms which underlies the renormalization
of the one-body single-$\beta$ and the two-body \zbb~decay operators
are different.

In closing, we address the role of many-body electroweak currents
in the redefinition of single-$\beta$-decay and \zbb-decay operators.
Within the shell model, one can use nuclear potentials derived within
chiral perturbation theory \cite{Fukui18,Ma19}, and include also 
the contributions of chiral two-body electroweak currents.
Studies along these lines have been recently carried out in
Refs.\cite{Gysbers19,Menendez11,Wang18}, where the authors find
significant contributions from two-body axial currents in
$\beta$-decay.  Concurrently, a study reported in Ref.~\cite{Rho19a}
argues that many-body electroweak currents should play a negligible
role in standard GT transitions (namely, $\beta$- and \dbb~decays) due
to the ``chiral filter'' mechanism\cite{Rho19a}.
The chiral filter mechanism may be no longer valid for \zbb~decay
since the transferred momentum is $\sim 100$ MeV, which will require
further investigations to fully understand the role of many-body
currents in SM calculations of \nmes.

\subsection{Comparison between SM calculations}\label{comparison}
\begin{figure}[H]
\begin{center}
\includegraphics[scale=0.6,angle=0]{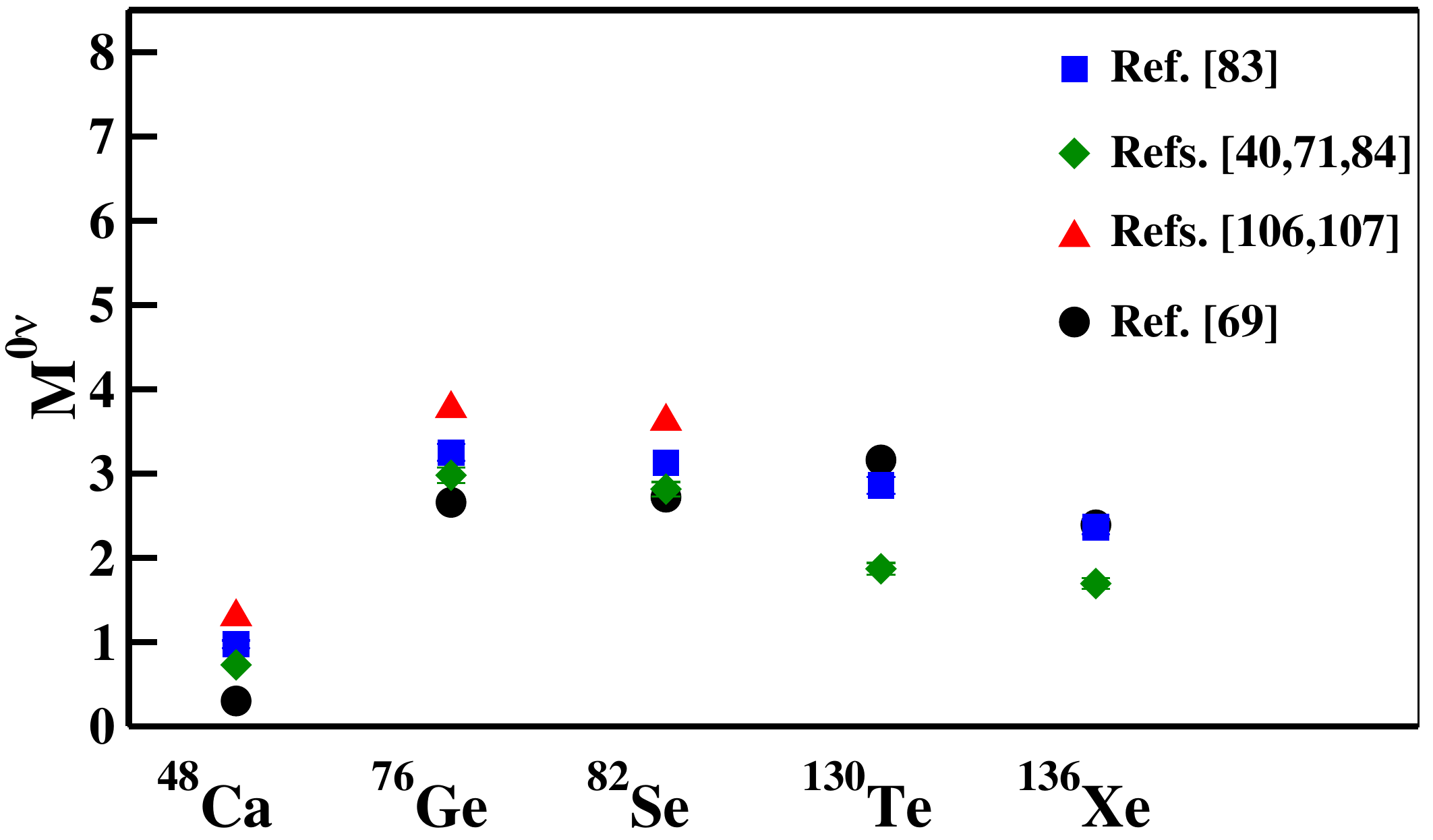}
\end{center}
\caption{\nme's calculated using different \heff~and \zbb-decay
  operators. Blue squares represent the results based on the
  Strasbourg-Madrid \heff's~\cite{Menendez18}. Green diamonds
  correspond to the calculations by Horoi and
  coworkers~\cite{Neacsu15,Senkov13,Senkov14}. 
  Results by Holt and Engel are indicated by the red
  triangles~\cite{Holt13d, Kwiatkowski14}.  
  Black dots indicate results by some of the present authors reported
  in Ref.~\cite{Coraggio20a}.}
\label{nmecomp}
\end{figure}

In Fig.~\ref{nmecomp}, we group most of the SM results for
$^{48}$Ca$\rightarrow$ $^{48}$Ti, $^{76}$Ge  $\rightarrow$ $^{76}$Se,
$^{82}$Se $\rightarrow$ $^{82}$Kr, $^{130}$Te $\rightarrow$
$^{130}$Xe, and $^{136}$Xe  $\rightarrow$ $^{136}$Ba.
We have chosen the results according to the following criteria:

\begin{enumerate}
\item[a)] All the SM calculations for a given transition are based on
  the same model space.
\item[b)] All the calculations use the closure approximation.
\item[c)] Whenever the calculations use different choices of SRCs, the
  average value and associated error bar is reported.
\end{enumerate}

The scale on the y-axis is consistent with the one employed in Fig. 5
of Ref.~\cite{Engel17}.
We stress that our criteria rule out, for the sake of consistency,
results from SM calculations where alternative approaches have been
followed.
For example, we recall that Horoi and coworkers have extensively
performed calculations beyond the closure
approximation~\cite{Senkov13,Senkov14,Senkov16}.
In particular, as already mentioned several times, results for the
\nme~ of $^{48}$Ca calculated with and without the closure
approximation differ by $\sim 10\%$.
Likewise, we neglect the results of large-scale SM calculations, where
model spaces larger than a single major shell are used.
This is the case, for instance, of the work by Horoi and Brown
\cite{Horoi13b} and by  Iwata and coworkers \cite{Iwata16}. In the
former, it was shown how the enlargement of the standard
$g_{9/2}dsh_{9/2}$ model space by the inclusion of the spin-orbit
partners of g$_{9/2}$ and h$_{9/2}$ orbitals leads to a 10-30\%
reduction of the calculated \nme for the $^{136}$Xe emitter.
In the latter, the authors reported the results for the \nme of $^{48}$Ca
based on large-scale shell-model calculations including two harmonic
oscillator shells ($sd$ and $pf$ shells). They found that \nme is
enhanced by about 30\% with respect to $pf$-shell calculations  when 
excitations up to $2\hbar \omega$ are explicitly included.

The spread among different results is rather narrow, except from the
$^{130}$Te $\rightarrow$ $^{130}$Xe \zbb~decay since the result in
Ref.~\cite{Neacsu15} are more than $30\%$ larger than those in
Refs.~\cite{Menendez18,Coraggio20a}.
We observe that the results from Refs.~\cite{Menendez18,Coraggio20a}
are close each other, and the \nme's~calculated by Holt and
Engel~\cite{Holt13d, Kwiatkowski14} are consistently larger than those
from other SM calculations.
The computational methods reported in the figure use substantially 
different SM effective Hamiltonians, yet they all lead to equally satisfactory
results for a large amount of structure data.
This leads us to argue that the SM approach is reliable for the study
of \zbb~decay.

Finally, comparing Fig.~\ref{nmecomp} with the compilation of results
reported in Fig. 5 in Ref.~\cite{Engel17}, we confirm that also these
more recent SM calculations provide values that are smaller than those
obtained with other nuclear models such as the Interacting Boson
Model, the QRPA, and Energy Density Functional methods.
Since the advantage of the nuclear shell model with respect to other
approaches is to include a larger number of nuclear correlations, one
may argue that by enlarging the dimension of the Hilbert space of
nuclear configurations it should be expected a reduction in magnitude
of the predicted values of \nme's, as it is found {\it e.g.}, in
Ref.~\cite{Horoi13b}.

\section{Conclusions}\label{conclusions}
In this paper, we have briefly reviewed  the present status of SM
calculations of \zbb-decay nuclear matrix elements.
More precisely, we have focused our attention to the
$^{48}$Ca$\rightarrow^{48}$Ti, $^{76}$Ge$\rightarrow^{76}$Se,
$^{82}$Se$\rightarrow^{82}$Kr, $^{130}$Te$\rightarrow^{130}$Xe, and
$^{136}$Xe$\rightarrow^{136}$Ba decays.
These transitions are relevant to the current and planned experimental
programs. 
We have considered calculations performed with phenomenological SM
Hamiltonians, as well as studies where SP energies, two-body matrix
elements, and effective decay operators have been derived from
realistic $V^{NN}$ potentials.
For completeness, important aspects that characterize the calculation
of \nme, such as the role of short-range correlations and the closure
approximation, have been briefly discussed.
These and other approximations, such as the size of the model space,
may affect the results by $\lesssim 30\%$.

We showed how different SM calculations, notwithstanding the diversity
of the effective Hamiltonians that are employed to calculate the
nuclear wave functions, exhibit a rather narrow spread among the
predictions of the nuclear matrix elements, making the  SM a solid and
reliable framework for \nme~ calculations.

\externalbibliography{yes}
\bibliography{biblio}

\begin{thebibliography}{-------}
\providecommand{\natexlab}[1]{#1}

\bibitem[Schechter and Valle(1982)]{Schechter:1981bd}
Schechter, J.; Valle, J.W.F.
\newblock {\em Phys. Rev. D} {\bf 1982}, {\em 25},~2951.

\bibitem[Davidson \em{et~al.}(2008)Davidson, Nardi, and Nir]{Davidson:2008bu}
Davidson, S.; Nardi, E.; Nir, Y.
\newblock {\em Phys. Rept.} {\bf 2008}, {\em 466},~105--177.

\bibitem[Gando \em{et~al.}(2013)Gando et~al.]{Gando:2012zm}
Gando, A.; others.
\newblock {\em Phys. Rev. Lett.} {\bf 2013}, {\em 110},~062502.

\bibitem[Agostini \em{et~al.}(2013)Agostini et~al.]{Agostini:2013mzu}
Agostini, M.; others.
\newblock {\em Phys. Rev. Lett.} {\bf 2013}, {\em 111},~122503.

\bibitem[Albert \em{et~al.}(2014)Albert et~al.]{Albert:2014awa}
Albert, J.B.; others.
\newblock {\em Nature} {\bf 2014}, {\em 510},~229--234.

\bibitem[Andringa \em{et~al.}(2016)Andringa et~al.]{Andringa:2015tza}
Andringa, S.; others.
\newblock {\em Adv. High Energy Phys.} {\bf 2016}, {\em 2016},~6194250.

\bibitem[Gando \em{et~al.}(2016)Gando et~al.]{KamLAND-Zen:2016pfg}
Gando, A.; others.
\newblock {\em Phys. Rev. Lett.} {\bf 2016}, {\em 117},~082503.

\bibitem[Elliott \em{et~al.}(2017)Elliott et~al.]{Elliott:2016ble}
Elliott, S.R.; others.
\newblock {\em J. Phys. Conf. Ser.} {\bf 2017}, {\em 888},~012035.

\bibitem[Agostini \em{et~al.}(2017)Agostini et~al.]{Agostini:2017iyd}
Agostini, M.; others.
\newblock {\em Nature} {\bf 2017}, {\em 544},~47.

\bibitem[Aalseth \em{et~al.}(2018)Aalseth et~al.]{Aalseth:2017btx}
Aalseth, C.E.; others.
\newblock {\em Phys. Rev. Lett.} {\bf 2018}, {\em 120},~132502.

\bibitem[Albert \em{et~al.}(2018)Albert et~al.]{Albert:2017owj}
Albert, J.B.; others.
\newblock {\em Phys. Rev. Lett.} {\bf 2018}, {\em 120},~072701.

\bibitem[Alduino \em{et~al.}(2018)Alduino et~al.]{Alduino:2017ehq}
Alduino, C.; others.
\newblock {\em Phys. Rev. Lett.} {\bf 2018}, {\em 120},~132501.

\bibitem[Agostini \em{et~al.}(2018)Agostini et~al.]{Agostini:2018tnm}
Agostini, M.; others.
\newblock {\em Phys. Rev. Lett.} {\bf 2018}, {\em 120},~132503.

\bibitem[Azzolini \em{et~al.}(2018)Azzolini et~al.]{Azzolini:2018dyb}
Azzolini, O.; others.
\newblock {\em Phys. Rev. Lett.} {\bf 2018}, {\em 120},~232502.

\bibitem[Kotila and Iachello(2012)]{Kotila12}
Kotila, J.; Iachello, F.
\newblock {\em Phys. Rev. C} {\bf 2012}, {\em 85},~034316.

\bibitem[Kotila and Iachello(2013)]{Kotila13}
Kotila, J.; Iachello, F.
\newblock {\em Phys. Rev. C} {\bf 2013}, {\em 87},~024313.

\bibitem[Engel and Men{\'e}ndez(2017)]{Engel17}
Engel, J.; Men{\'e}ndez, J.
\newblock {\em Rep. Prog. Phys.} {\bf 2017}, {\em 80},~046301.

\bibitem[Pastore \em{et~al.}(2018)Pastore, Carlson, Cirigliano, Dekens,
  Mereghetti, and Wiringa]{Pastore18}
Pastore, S.; Carlson, J.; Cirigliano, V.; Dekens, W.; Mereghetti, E.; Wiringa,
  R.B.
\newblock {\em Phys. Rev. C} {\bf 2018}, {\em 97},~014606.

\bibitem[Wang \em{et~al.}(2019)Wang, Hayes, Carlson, Dong, Mereghetti, Pastore,
  and Wiringa]{Wang:2019hjy}
Wang, X.; Hayes, A.; Carlson, J.; Dong, G.; Mereghetti, E.; Pastore, S.;
  Wiringa, R.
\newblock {\em Phys. Lett. B} {\bf 2019}, {\em 798},~134974.

\bibitem[Basili \em{et~al.}(2020)Basili, Yao, Engel, Hergert, Lockner, Maris,
  and Vary]{Basili:2019gvn}
Basili, R.; Yao, J.; Engel, J.; Hergert, H.; Lockner, M.; Maris, P.; Vary, J.
\newblock {\em Phys. Rev. C} {\bf 2020}, {\em 102},~014302.

\bibitem[Hergert \em{et~al.}(2016)Hergert, Bogner, Morris, Schwenk, and
  Tsukiyama]{Hergert16}
Hergert, H.; Bogner, S.K.; Morris, T.D.; Schwenk, A.; Tsukiyama, K.
\newblock {\em Phys. Rep.} {\bf 2016}, {\em 621},~165.

\bibitem[Griffin and Wheeler(1957)]{Griffin57}
Griffin, J.J.; Wheeler, J.A.
\newblock {\em Phys. Rev.} {\bf 1957}, {\em 108},~311--327.

\bibitem[Barea and Iachello(2009)]{Barea09}
Barea, J.; Iachello, F.
\newblock {\em Phys. Rev. C} {\bf 2009}, {\em 79},~044301.

\bibitem[Barea \em{et~al.}(2012)Barea, Kotila, and Iachello]{Barea12}
Barea, J.; Kotila, J.; Iachello, F.
\newblock {\em Phys. Rev. Lett.} {\bf 2012}, {\em 109},~042501.

\bibitem[Barea \em{et~al.}(2013)Barea, Kotila, and Iachello]{Barea13}
Barea, J.; Kotila, J.; Iachello, F.
\newblock {\em Phys. Rev. C} {\bf 2013}, {\em 87},~014315.

\bibitem[Sim(2009)]{Simkovic09}
{\em Phys. Rev. C} {\bf 2009}, {\em 79},~055501.

\bibitem[Fang \em{et~al.}(2011)Fang, Faessler, Rodin, and
  \ifmmode~\check{S}\else \v{S}\fi{}imkovic]{Fang11}
Fang, D.L.; Faessler, A.; Rodin, V.; \ifmmode~\check{S}\else \v{S}\fi{}imkovic,
  F.
\newblock {\em Phys. Rev. C} {\bf 2011}, {\em 83},~034320.

\bibitem[Faessler \em{et~al.}(2012)Faessler, Rodin, and Simkovic]{Faessler12}
Faessler, A.; Rodin, V.; Simkovic, F.
\newblock {\em J. Phys. G} {\bf 2012}, {\em 39},~124006.

\bibitem[Rodr\'{\i}guez and Mart\'{\i}nez-Pinedo(2010)]{Rodriguez10}
Rodr\'{\i}guez, T.R.; Mart\'{\i}nez-Pinedo, G.
\newblock {\em Phys. Rev. Lett.} {\bf 2010}, {\em 105},~252503.

\bibitem[Song \em{et~al.}(2014)Song, Yao, Ring, and Meng]{Song14}
Song, L.S.; Yao, J.M.; Ring, P.; Meng, J.
\newblock {\em Phys. Rev. C} {\bf 2014}, {\em 90},~054309.

\bibitem[Yao \em{et~al.}(2015)Yao, Song, Hagino, Ring, and Meng]{Yao15}
Yao, J.M.; Song, L.S.; Hagino, K.; Ring, P.; Meng, J.
\newblock {\em Phys. Rev. C} {\bf 2015}, {\em 91},~024316.

\bibitem[Song \em{et~al.}(2017)Song, Yao, Ring, and Meng]{Song17}
Song, L.S.; Yao, J.M.; Ring, P.; Meng, J.
\newblock {\em Phys. Rev. C} {\bf 2017}, {\em 95},~024305.

\bibitem[Jiao \em{et~al.}(2017)Jiao, Engel, and Holt]{Jiao17}
Jiao, C.F.; Engel, J.; Holt, J.D.
\newblock {\em Phys. Rev. C} {\bf 2017}, {\em 96},~054310.

\bibitem[Yao \em{et~al.}(2018)Yao, Engel, Wang, Jiao, and Hergert]{Yao18}
Yao, J.M.; Engel, J.; Wang, L.J.; Jiao, C.F.; Hergert, H.
\newblock {\em Phys. Rev. C} {\bf 2018}, {\em 98},~054311.

\bibitem[Jiao \em{et~al.}(2018)Jiao, Horoi, and Neacsu]{Jiao18}
Jiao, C.F.; Horoi, M.; Neacsu, A.
\newblock {\em Phys. Rev. C} {\bf 2018}, {\em 98},~064324.

\bibitem[Jiao and Johnson(2019)]{Jiao19}
Jiao, C.; Johnson, C.W.
\newblock {\em Phys. Rev. C} {\bf 2019}, {\em 100},~031303.

\bibitem[Men\'endez \em{et~al.}(2009{\natexlab{a}})Men\'endez, Poves, Caurier,
  and Nowacki]{Menendez09a}
Men\'endez, J.; Poves, A.; Caurier, E.; Nowacki, F.
\newblock {\em Phys. Rev. C} {\bf 2009}, {\em 80},~048501.

\bibitem[Men\'endez \em{et~al.}(2009{\natexlab{b}})Men\'endez, Poves, Caurier,
  and Nowacki]{Menendez09b}
Men\'endez, J.; Poves, A.; Caurier, E.; Nowacki, F.
\newblock {\em Nucl. Phys. A} {\bf 2009}, {\em 818},~139.

\bibitem[Horoi and Brown(2013)]{Horoi13b}
Horoi, M.; Brown, B.A.
\newblock {\em Phys. Rev. Lett.} {\bf 2013}, {\em 110},~222502.

\bibitem[Neacsu and Horoi(2015)]{Neacsu15}
Neacsu, A.; Horoi, M.
\newblock {\em Phys. Rev. C} {\bf 2015}, {\em 91},~024309.

\bibitem[Brown \em{et~al.}(2015)Brown, Fang, and Horoi]{Brown15}
Brown, B.A.; Fang, D.L.; Horoi, M.
\newblock {\em Phys. Rev. C} {\bf 2015}, {\em 92},~041301.

\bibitem[Tanabashi and {\it et al}(2018)]{PDG18}
Tanabashi, M.; {\it et al}.
\newblock {\em Phys. Rev. D} {\bf 2018}, {\em 98},~030001.

\bibitem[Towner(1987)]{Towner87}
Towner, I.S.
\newblock {\em Phys. Rep.} {\bf 1987}, {\em 155},~263.

\bibitem[Chou \em{et~al.}(1993)Chou, Warburton, and Brown]{Chou:1993zz}
Chou, W.T.; Warburton, E.K.; Brown, B.A.
\newblock {\em Phys. Rev. C} {\bf 1993}, {\em 47},~163--177.

\bibitem[Suhonen(2017{\natexlab{a}})]{Suhonen17b}
Suhonen, J.T.
\newblock {\em Frontiers in Physics} {\bf 2017}, {\em 5},~55.

\bibitem[Suhonen(2017{\natexlab{b}})]{Suhonen17a}
Suhonen, J.
\newblock {\em Phys. Rev. C} {\bf 2017}, {\em 96},~055501.

\bibitem[Park \em{et~al.}(1993)Park, Min, and Rho]{Park93}
Park, T.S.; Min, D.P.; Rho, M.
\newblock {\em Phys. Rep.} {\bf 1993}, {\em 233},~341.

\bibitem[Park \em{et~al.}(1996)Park, Min, and Rho]{Park:1995pn}
Park, T.S.; Min, D.P.; Rho, M.
\newblock {\em Nucl. Phys.} {\bf 1996}, {\em A596},~515--552.

\bibitem[Baroni \em{et~al.}(2016)Baroni, Girlanda, Pastore, Schiavilla, and
  Viviani]{Baroni16b}
Baroni, A.; Girlanda, L.; Pastore, S.; Schiavilla, R.; Viviani, M.
\newblock {\em Phys. Rev. C} {\bf 2016}, {\em 93},~015501.

\bibitem[Krebs \em{et~al.}(2017)Krebs, Epelbaum, and Meißner]{Krebs:2016rqz}
Krebs, H.; Epelbaum, E.; Meißner, U.G.
\newblock {\em Annals Phys.} {\bf 2017}, {\em 378},~317--395.

\bibitem[Krebs \em{et~al.}(2020)Krebs, Epelbaum, and
  Mei\ss{}ner]{Krebs:2020rms}
Krebs, H.; Epelbaum, E.; Mei\ss{}ner, U.G.
\newblock {\em Phys. Rev. C} {\bf 2020}, {\em 101},~055502.

\bibitem[Pastore \em{et~al.}(2018)Pastore, Baroni, Carlson, Gandolfi, Pieper,
  Schiavilla, and Wiringa]{Pastore:2017uwc}
Pastore, S.; Baroni, A.; Carlson, J.; Gandolfi, S.; Pieper, S.C.; Schiavilla,
  R.; Wiringa, R.
\newblock {\em Phys. Rev. C} {\bf 2018}, {\em 97},~022501.

\bibitem[King \em{et~al.}(2020)King, Andreoli, Pastore, Piarulli, Schiavilla,
  Wiringa, Carlson, and Gandolfi]{King:2020wmp}
King, G.B.; Andreoli, L.; Pastore, S.; Piarulli, M.; Schiavilla, R.; Wiringa,
  R.B.; Carlson, J.; Gandolfi, S.
\newblock {\em Phys. Rev. C} {\bf 2020}, {\em 102},~025501.

\bibitem[Gysbers \em{et~al.}(2019)Gysbers, Hagen, Holt, Jansen, Morris,
  Navr{\'a}til, Papenbrock, Quaglioni, Schwenk, Stroberg, and Wendt]{Gysbers19}
Gysbers, P.; Hagen, G.; Holt, J.D.; Jansen, G.R.; Morris, T.D.; Navr{\'a}til,
  P.; Papenbrock, T.; Quaglioni, S.; Schwenk, A.; Stroberg, S.R.; Wendt, K.A.
\newblock {\em Nature Phys.} {\bf 2019}, {\em 15},~428.

\bibitem[Mayer(1949)]{Mayer49}
Mayer, M.G.
\newblock {\em Phys. Rev.} {\bf 1949}, {\em 75},~1969.

\bibitem[Haxel \em{et~al.}(1949)Haxel, Jensen, and Suess]{Haxel49}
Haxel, O.; Jensen, J.H.D.; Suess, H.E.
\newblock {\em Phys. Rev.} {\bf 1949}, {\em 75},~1766.

\bibitem[Mayer and Jensen(1955)]{Mayer55}
Mayer, M.G.; Jensen, J.H.D.
\newblock {\em Elementary Theory of Nuclear Shell Structure}; John Wiley, New
  York,  1955.

\bibitem[Machleidt(2017)]{Machleidt:2017vls}
Machleidt, R.
\newblock {\em Int. J. Mod. Phys. E} {\bf 2017}, {\em 26},~1730005.

\bibitem[Elliott(1969)]{Elliott69a}
Elliott, J.P.
\newblock Nuclear forces and the structure of nuclei.
\newblock  Carg\`ese Lectures in Physics, Vol. 3; Jean, M., Ed. Gordon and
  Breach, New York,  1969, p. 337.

\bibitem[Talmi(2003)]{Talmi03}
Talmi, I.
\newblock {\em Adv. Nucl. Phys.} {\bf 2003}, {\em 27},~1.

\bibitem[Caurier \em{et~al.}(2005)Caurier, Mart\'{\i}nez-Pinedo, Nowacki,
  Poves, and Zuker]{Caurier05}
Caurier, E.; Mart\'{\i}nez-Pinedo, G.; Nowacki, F.; Poves, A.; Zuker, A.P.
\newblock {\em Rev. Mod. Phys.} {\bf 2005}, {\em 77},~427--488.

\bibitem[Bethe(1971)]{Bethe71}
Bethe, H.A.
\newblock {\em Annu. Rev. Nucl. Sci.} {\bf 1971}, {\em 21},~93.

\bibitem[Kortelainen \em{et~al.}(2007)Kortelainen, Civitarese, Suhonen, and
  Toivanen]{Kortelainen07}
Kortelainen, M.; Civitarese, O.; Suhonen, J.; Toivanen, J.
\newblock {\em Phys. Lett. B} {\bf 2007}, {\em 647},~128.

\bibitem[Wu \em{et~al.}(1985)Wu, Song, Kuo, Cheng, and Strottman]{Wu85}
Wu, H.F.; Song, H.Q.; Kuo, T.T.S.; Cheng, W.K.; Strottman, D.
\newblock {\em Phys. Lett. B} {\bf 1985}, {\em 162},~227.

\bibitem[Miller and Spencer(1976)]{Miller76}
Miller, G.A.; Spencer, J.E.
\newblock {\em Ann. Phys. (NY)} {\bf 1976}, {\em 100},~562.

\bibitem[Neacsu \em{et~al.}(2012)Neacsu, Stoica, and Horoi]{Neacsu12}
Neacsu, A.; Stoica, S.; Horoi, M.
\newblock {\em Phys. Rev. C} {\bf 2012}, {\em 86},~067304.

\bibitem[Feldmeier \em{et~al.}(1998)Feldmeier, Neff, Roth, and
  Schnack]{Feldmeier98}
Feldmeier, H.; Neff, T.; Roth, R.; Schnack, J.
\newblock {\em Nucl. Phys. A} {\bf 1998}, {\em 632},~61.

\bibitem[Coraggio \em{et~al.}(2019)Coraggio, Itaco, and Mancino]{Coraggio19b}
Coraggio, L.; Itaco, N.; Mancino, R.
\newblock {\em arXiv:1910.04146 [nucl-th]} {\bf 2019}.
\newblock to be published in the Conference Proceedings of the 27th
  International Nuclear Physics Conference, 29 July - 2 August 2019, Glasgow
  (UK).

\bibitem[Coraggio \em{et~al.}(2020)Coraggio, Gargano, Itaco, Mancino, and
  Nowacki]{Coraggio20a}
Coraggio, L.; Gargano, A.; Itaco, N.; Mancino, R.; Nowacki, F.
\newblock {\em Phys. Rev. C} {\bf 2020}, {\em 101},~044315.

\bibitem[Bogner \em{et~al.}(2002)Bogner, Kuo, Coraggio, Covello, and
  Itaco]{Bogner02}
Bogner, S.; Kuo, T.T.S.; Coraggio, L.; Covello, A.; Itaco, N.
\newblock {\em Phys. Rev. C} {\bf 2002}, {\em 65},~051301(R).

\bibitem[Sen'kov and Horoi(2013)]{Senkov13}
Sen'kov, R.A.; Horoi, M.
\newblock {\em Phys. Rev. C} {\bf 2013}, {\em 88},~064312.

\bibitem[\ifmmode~\check{S}\else \v{S}\fi{}imkovic
  \em{et~al.}(2008)\ifmmode~\check{S}\else \v{S}\fi{}imkovic, Faessler, Rodin,
  Vogel, and Engel]{Simkovic08}
\ifmmode~\check{S}\else \v{S}\fi{}imkovic, F.; Faessler, A.; Rodin, V.; Vogel,
  P.; Engel, J.
\newblock {\em Phys. Rev. C} {\bf 2008}, {\em 77},~045503.

\bibitem[Haxton and Stephenson~Jr.(1984)]{Haxton84}
Haxton, W.C.; Stephenson~Jr., G.J.
\newblock {\em Prog. Part. Nucl. Phys.} {\bf 1984}, {\em 12},~409.

\bibitem[Tomoda(1991)]{Tomoda91}
Tomoda, T.
\newblock {\em Rep. Prog. Phys.} {\bf 1991}, {\em 54},~53.

\bibitem[Elliott and Petr(2002)]{Elliott02}
Elliott, S.R.; Petr, V.
\newblock {\em Annu. Rev. Nucl. Part. Sci.} {\bf 2002}, {\em 52},~115--151.

\bibitem[Neacsu and Stoica(2013)]{Neacsu13}
Neacsu, A.; Stoica, S.
\newblock {\em J. Phys. G} {\bf 2013}, {\em 41},~015201.

\bibitem[Ikeda(1964)]{Ikeda64}
Ikeda, K.
\newblock {\em Prog. Theor. Phys.} {\bf 1964}, {\em 31},~434.

\bibitem[Honma \em{et~al.}(2004)Honma, Otsuka, Brown, and Mizusaki]{Honma04}
Honma, M.; Otsuka, T.; Brown, B.A.; Mizusaki, T.
\newblock {\em Phys. Rev. C} {\bf 2004}, {\em 69},~034335.

\bibitem[Honma \em{et~al.}(2005)Honma, Otsuka, Brown, and Mizusaki]{Honma05}
Honma, M.; Otsuka, T.; Brown, B.A.; Mizusaki, T.
\newblock {\em Eur. Phys. J. A} {\bf 2005}, {\em 25,~s01},~499.

\bibitem[Poves and Zuker(1981)]{Poves81}
Poves, A.; Zuker, A.
\newblock {\em Phys. Rep.} {\bf 1981}, {\em 70},~235.

\bibitem[Poves \em{et~al.}(2001)Poves, S{\'a}nchez-Solano, Caurier, and
  Nowacki]{Poves01}
Poves, A.; S{\'a}nchez-Solano, J.; Caurier, E.; Nowacki, F.
\newblock {\em Nucl. Phys. A} {\bf 2001}, {\em 694},~157.

\bibitem[Richter \em{et~al.}(1991)Richter, der Merwe, Julies, and
  Brown]{Richter91}
Richter, W.A.; der Merwe, M.G.V.; Julies, R.E.; Brown, B.A.
\newblock {\em Nucl. Phys. A} {\bf 1991}, {\em 523},~325.

\bibitem[Menendez(2018)]{Menendez18}
Menendez, J.
\newblock {\em J. Phys. G} {\bf 2018}, {\em 45},~014003.

\bibitem[Sen'kov \em{et~al.}(2014)Sen'kov, Horoi, and Brown]{Senkov14}
Sen'kov, R.A.; Horoi, M.; Brown, B.A.
\newblock {\em Phys. Rev. C} {\bf 2014}, {\em 89},~054304.

\bibitem[Honma \em{et~al.}(2009)Honma, Otsuka, Mizusaki, and
  Hjorth-Jensen]{Honma09}
Honma, M.; Otsuka, T.; Mizusaki, T.; Hjorth-Jensen, M.
\newblock {\em Phys. Rev. C} {\bf 2009}, {\em 80},~064323.

\bibitem[Qi and Xu(2012)]{Qi12}
Qi, C.; Xu, Z.X.
\newblock {\em Phys. Rev. C} {\bf 2012}, {\em 86},~044323.

\bibitem[Sen'kov and Horoi(2016)]{Senkov16}
Sen'kov, R.A.; Horoi, M.
\newblock {\em Phys. Rev. C} {\bf 2016}, {\em 93},~044334.

\bibitem[Alexander(2020)]{Barabash20}
Alexander, B.
\newblock {\em Universe} {\bf 2020}, {\em 6},~159.

\bibitem[Kostensalo and Suhonen(2020)]{Kostensalo20}
Kostensalo, J.; Suhonen, J.
\newblock {\em Phys. Lett. B} {\bf 2020}, {\em 802},~135192.

\bibitem[Caurier \em{et~al.}(2012)Caurier, Nowacki, and Poves]{Caurier12}
Caurier, E.; Nowacki, F.; Poves, A.
\newblock {\em Phys. Lett. B} {\bf 2012}, {\em 711},~62.

\bibitem[Brandow(1967)]{Brandow67}
Brandow, B.H.
\newblock {\em Rev. Mod. Phys.} {\bf 1967}, {\em 39},~771.

\bibitem[Kuo \em{et~al.}(1971)Kuo, Lee, and Ratcliff]{Kuo71}
Kuo, T.T.S.; Lee, S.Y.; Ratcliff, K.F.
\newblock {\em Nucl. Phys. A} {\bf 1971}, {\em 176},~65.

\bibitem[Krenciglowa and Kuo(1974)]{Krenciglowa74}
Krenciglowa, E.M.; Kuo, T.T.S.
\newblock {\em Nucl. Phys. A} {\bf 1974}, {\em 235},~171.

\bibitem[Lee and Suzuki(1980)]{Lee80}
Lee, S.Y.; Suzuki, K.
\newblock {\em Phys. Lett. B} {\bf 1980}, {\em 91},~173.

\bibitem[Suzuki and Lee(1980)]{Suzuki80}
Suzuki, K.; Lee, S.Y.
\newblock {\em Prog. Theor. Phys.} {\bf 1980}, {\em 64},~2091.

\bibitem[Coraggio \em{et~al.}(2012)Coraggio, Covello, Gargano, Itaco, and
  Kuo]{Coraggio12a}
Coraggio, L.; Covello, A.; Gargano, A.; Itaco, N.; Kuo, T.T.S.
\newblock {\em Ann. Phys. (NY)} {\bf 2012}, {\em 327},~2125.

\bibitem[Stroberg \em{et~al.}(2019)Stroberg, Hergert, Bogner, and
  Holt]{Stroberg19}
Stroberg, S.R.; Hergert, H.; Bogner, S.K.; Holt, J.D.
\newblock {\em Annu. Rev. Nucl. Part. Sci.} {\bf 2019}, {\em 69},~307--362.

\bibitem[Ellis and Osnes(1977)]{Ellis77}
Ellis, P.J.; Osnes, E.
\newblock {\em Rev. Mod. Phys.} {\bf 1977}, {\em 49},~777.

\bibitem[Suzuki and Okamoto(1995)]{Suzuki95}
Suzuki, K.; Okamoto, R.
\newblock {\em Prog. Theor. Phys.} {\bf 1995}, {\em 93},~905.

\bibitem[Towner and Khanna(1983)]{Towner83}
Towner, I.S.; Khanna, K.F.C.
\newblock {\em Nucl. Phys. A} {\bf 1983}, {\em 399},~334.

\bibitem[Lacombe \em{et~al.}(1980)Lacombe, Loiseau, Richard, Mau, C\^ot\`e,
  Pir\'es, and Tourreil]{Lacombe80}
Lacombe, M.; Loiseau, B.; Richard, J.M.; Mau, R.V.; C\^ot\`e, J.; Pir\'es, P.;
  Tourreil, R.D.
\newblock {\em Phys. Rev C} {\bf 1980}, {\em 21},~861.

\bibitem[Reid(1968)]{Reid68}
Reid, R.V.
\newblock {\em Ann. Phys. (N.Y.)} {\bf 1968}, {\em 50},~411.

\bibitem[Krenciglowa \em{et~al.}(1976)Krenciglowa, Kung, Kuo, and
  Osnes]{Krenciglowa76}
Krenciglowa, E.M.; Kung, C.L.; Kuo, T.T.S.; Osnes, E.
\newblock {\em Ann. Phys.} {\bf 1976}, {\em 101},~154.

\bibitem[Kuo and Brown(1966)]{Kuo66}
Kuo, T.T.S.; Brown, G.E.
\newblock {\em Nucl. Phys.} {\bf 1966}, {\em 85},~40.

\bibitem[Entem and Machleidt(2002)]{Entem02}
Entem, D.R.; Machleidt, R.
\newblock {\em Phys. Rev. C} {\bf 2002}, {\em 66},~014002.

\bibitem[Holt and Engel(2013)]{Holt13d}
Holt, J.D.; Engel, J.
\newblock {\em Phys. Rev. C} {\bf 2013}, {\em 87},~064315.

\bibitem[Kwiatkowski \em{et~al.}(2014)Kwiatkowski, Brunner, Holt, Chaudhuri,
  Chowdhury, Eibach, Engel, Gallant, Grossheim, Horoi, Lennarz, Macdonald,
  Pearson, Schultz, Simon, Senkov, Simon, Zuber, and Dilling]{Kwiatkowski14}
Kwiatkowski, A.A.; Brunner, T.; Holt, J.D.; Chaudhuri, A.; Chowdhury, U.;
  Eibach, M.; Engel, J.; Gallant, A.T.; Grossheim, A.; Horoi, M.; Lennarz, A.;
  Macdonald, T.D.; Pearson, M.R.; Schultz, B.E.; Simon, M.C.; Senkov, R.A.;
  Simon, V.V.; Zuber, K.; Dilling, J.
\newblock {\em Phys. Rev. C} {\bf 2014}, {\em 89},~045502.

\bibitem[Horoi(2013)]{Horoi13c}
Horoi, M.
\newblock {\em J. Phys. Conf. Ser.} {\bf 2013}, {\em 413},~012020.

\bibitem[Machleidt(2001)]{Machleidt01b}
Machleidt, R.
\newblock {\em Phys. Rev. C} {\bf 2001}, {\em 63},~024001.

\bibitem[Coraggio \em{et~al.}(2017)Coraggio, De~Angelis, Fukui, Gargano, and
  Itaco]{Coraggio17a}
Coraggio, L.; De~Angelis, L.; Fukui, T.; Gargano, A.; Itaco, N.
\newblock {\em Phys. Rev. C} {\bf 2017}, {\em 95},~064324.

\bibitem[Coraggio \em{et~al.}(2019)Coraggio, De~Angelis, Fukui, Gargano, Itaco,
  and Nowacki]{Coraggio19a}
Coraggio, L.; De~Angelis, L.; Fukui, T.; Gargano, A.; Itaco, N.; Nowacki, F.
\newblock {\em Phys. Rev. C} {\bf 2019}, {\em 100},~014316.

\bibitem[Pastore \em{et~al.}(2008)Pastore, Schiavilla, and
  Goity]{Pastore:2008ui}
Pastore, S.; Schiavilla, R.; Goity, J.L.
\newblock {\em Phys. Rev.} {\bf 2008}, {\em C78},~064002.

\bibitem[Pastore \em{et~al.}(2009)Pastore, Girlanda, Schiavilla, Viviani, and
  Wiringa]{Pastore09}
Pastore, S.; Girlanda, L.; Schiavilla, R.; Viviani, M.; Wiringa, R.B.
\newblock {\em Phys. Rev. C} {\bf 2009}, {\em 80},~034004.

\bibitem[Pastore \em{et~al.}(2011)Pastore, Girlanda, Schiavilla, and
  Viviani]{Pastore:2011ip}
Pastore, S.; Girlanda, L.; Schiavilla, R.; Viviani, M.
\newblock {\em Phys. Rev.} {\bf 2011}, {\em C84},~024001.

\bibitem[Pastore \em{et~al.}(2014)Pastore, Wiringa, Pieper, and
  Schiavilla]{Pastore:2014oda}
Pastore, S.; Wiringa, R.B.; Pieper, S.C.; Schiavilla, R.
\newblock {\em Phys. Rev.} {\bf 2014}, {\em C90},~024321.

\bibitem[Pastore \em{et~al.}(2013)Pastore, Pieper, Schiavilla, and
  Wiringa]{Pastore:2012rp}
Pastore, S.; Pieper, S.C.; Schiavilla, R.; Wiringa, R.B.
\newblock {\em Phys. Rev.} {\bf 2013}, {\em C87},~035503.

\bibitem[Piarulli \em{et~al.}(2013)Piarulli, Girlanda, Marcucci, Pastore,
  Schiavilla, and Viviani]{Piarulli13}
Piarulli, M.; Girlanda, L.; Marcucci, L.E.; Pastore, S.; Schiavilla, R.;
  Viviani, M.
\newblock {\em Phys. Rev. C} {\bf 2013}, {\em 87},~014006.

\bibitem[Kolling \em{et~al.}(2009)Kolling, Epelbaum, Krebs, and
  Meissner]{Kolling:2009iq}
Kolling, S.; Epelbaum, E.; Krebs, H.; Meissner, U.G.
\newblock {\em Phys. Rev.} {\bf 2009}, {\em C80},~045502.

\bibitem[Kolling \em{et~al.}(2011)Kolling, Epelbaum, Krebs, and
  Meissner]{Kolling:2011mt}
Kolling, S.; Epelbaum, E.; Krebs, H.; Meissner, U.G.
\newblock {\em Phys. Rev.} {\bf 2011}, {\em C84},~054008.

\bibitem[Bacca and Pastore(2014)]{Bacca:2014tla}
Bacca, S.; Pastore, S.
\newblock {\em J. Phys.} {\bf 2014}, {\em G41},~123002.

\bibitem[Fukui \em{et~al.}(2018)Fukui, De~Angelis, Ma, Coraggio, Gargano,
  Itaco, and Xu]{Fukui18}
Fukui, T.; De~Angelis, L.; Ma, Y.Z.; Coraggio, L.; Gargano, A.; Itaco, N.; Xu,
  F.R.
\newblock {\em Phys. Rev. C} {\bf 2018}, {\em 98},~044305.

\bibitem[Ma \em{et~al.}(2019)Ma, Coraggio, De~Angelis, Fukui, Gargano, Itaco,
  and Xu]{Ma19}
Ma, Y.Z.; Coraggio, L.; De~Angelis, L.; Fukui, T.; Gargano, A.; Itaco, N.; Xu,
  F.R.
\newblock {\em Phys. Rev. C} {\bf 2019}, {\em 100},~034324.

\bibitem[Men\'endez \em{et~al.}(2011)Men\'endez, Gazit, and
  Schwenk]{Menendez11}
Men\'endez, J.; Gazit, D.; Schwenk, A.
\newblock {\em Phys. Rev. Lett.} {\bf 2011}, {\em 107},~062501.

\bibitem[Wang \em{et~al.}(2018)Wang, Engel, and Yao]{Wang18}
Wang, L.J.; Engel, J.; Yao, J.M.
\newblock {\em Phys. Rev. C} {\bf 2018}, {\em 98},~031301.

\bibitem[Rho(2019)]{Rho19a}
Rho, M.
\newblock A Solution to the Quenched $g_A$ Problem in Nuclei and Dense Baryonic
  Matter.
\newblock {\em arXiv:1903.09976[nucl-th]} {\bf 2019}.

\bibitem[Iwata \em{et~al.}(2016)Iwata, Shimizu, Otsuka, Utsuno, Men\'endez,
  Honma, and Abe]{Iwata16}
Iwata, Y.; Shimizu, N.; Otsuka, T.; Utsuno, Y.; Men\'endez, J.; Honma, M.; Abe,
  T.
\newblock {\em Phys. Rev. Lett.} {\bf 2016}, {\em 116},~112502.

\end{thebibliography}

\end{document}